\documentstyle[11pt]{article}
\def\be{\begin{equation}}
\def\bear{\begin{array}}
\def\eear{\end{array}}
\def\bea{\begin{eqnarray}}
\def\eea{\end{eqnarray}}
\def\Br{\Biggr}
\def\br{\biggr}
\def\Bl{\Biggl}
\def\bl{\biggl}
\def\l{\label}
\def\r{\ref}

\def\La{\Lambda}

\def\ee{\end{equation}}

\def\c{\cite}
\begin{document}
\begin{center}
\LARGE {\bf Large-N limit of   non-local 2D generalized Yang-Mills
theories  on non-orientable surface}
\end{center}
\begin{center}
 {\bf Kh. Saaidi {\footnote {E-mail: ksaaidi@uok.ac.ir}}}\\
 {\it Department of Science, University of Kurdistan, Pasdaran Ave., Sanandaj, Iran} \\
 {\it Institute for Studies in Theoretical Physics and Mathemaics,
 P.O.Box, 19395-5531, Tehran, Iran}\\
\end{center}
\vskip 3cm
\begin{center}
{\bf{Abstract}}
\end{center}
The large-group behavior of the
 non-local two dimensional generalized Yang-Mills theories (nlgYM$_2$'s)
 on  arbitrary closed non-orientable surfaces is investigated. It is shown that
all order of $\phi^{2k}$ model of these theories have thired order
phase transition only on projective plane (RP$^2$).  Also  the
phase structure of $\phi^2 + \frac{\gamma}{4}\phi^4$ model of
nlgYM$_2$ is studied and it is found  that for $\gamma >0$, this
model has third order phase transition  only on RP$^2$ and for
$\gamma<0$ it has third  order phase transition on any closed
non-orientable surfaces
 except RP$^2$ and Klein bottel.

\newpage
{\section{Introduction}}

 In recent years there have been much
effort  to analyze the different aspects of
two dimensional Yang-Mills (YM$_2$) theory [1-9]. The
YM$_2$  theory is defined by the Lagrangian ${\rm
tr}(\frac{1}{4}F^2)$ on a Riemann surface, where $F$ is the field
strength  tensor. This theory is equivalent to the so-called $\bf
B $- $\bf F$ theory, which characterized by the Lagrangian $i{\rm
tr }(BF) + {\rm tr}(B^2)$, such as invariance
  under area preserving diffeomorphisms and lack of any
propagating degrees of freedom \c{b1}. These properties are also
shared by a large  class of theories, called the generalized two
dimensional  Yang-Mills (gYM$_2$'s) theories. These theories,
however, are defined  by replacing  an arbitrary class function of
$B$ instead  of tr$(B^2)$ \c{e1}. Several aspect of this theories
such as, partition function, generating functional and large -N
limit  on an arbitrary two dimensional  orientable  and
non-orientable surfaces has been discussed in [16-22]. There is
another way to generalize YM$_2$ and gYM$_2$ and that is to use a
nonlocal action for the auxiliary field, leading to the so-called
nonlocal YM$_2$ (nlYM$_2$) and nonlocal gYM$_2$(nlgYM$_2$)
theories, respectively \c{kh1}.  Several aspects of nlYM$_2$
nlgYM$_2$, such as, classical behavior, wave function, partition
function, generating functional, and also large- $N$ limit of it,
have been studied on orientable surfaces in [11-15]. In all of
these theories, the solution  appear as some infinite summations
over the irreducible representations of the gauge group. In the
large - N limit, however, these summations are replaced by
suitable path integrals over continuous parameters characterizing
the Young tableaux, and saddle-point analysis shows that the only
significant representation is the classical one, which minimizes
some effective action. This continuous parameters characterizing
the representation  is a constrained, as the length of the rows of
the Young tableau is non-increasing. So for small values of the
surface area, the classical solution satisfies the constraint; for
large values of the surface area, it dose not. Therefore  the
dominating representation is not  the one, which minimizes the
effective  action. This introduces a phase transition  between
these  two regime. Some problem has been studied for special cases
of YM$_2$[17, 23], gYM$_2$ [20, 21, 22], nlYM$_2$[12, 14],
nlgYM$_2$ in [13, 15] on sphere. In this  paper I investigate this
problem ( large-N limit)  of nlgYM$_2$ theories on arbitrary
non-orientable surface.

The scheme of this paper is the following. In sect. 2, I briefly
review the large-N limit of nlgYM$_2$ theories of $U(N)$ gauge
group on any arbitrary  non-orientable surface and obtain
effective action of theory at large-N limit. In sect. 3, I study
the phase structure of nlgYM$_2$ for $\phi^{2k}$ model in all
order  on an arbitrary surface. In sect. 4,  I study the phase
structure of theory for $\phi^2 + \frac{\gamma}{4} \phi ^4$
models on any arbitrary non-orientable surface.

{\section{Preliminaries}}

In \c{kh1}, a non-local generalized two dimensional Yang-Mills
(nlgYM$_2$'s) theories was defined as:

\be \l{1}
 e^S:= \int DB \exp {\Br \{} \int i{\rm tr}(BF) d\mu +
\omega {\br [}\int \Lambda(B) d\mu {\bl]}{\Bl \}},
 \ee
where B is an auxiliary field at the adjoint representation of
gauge group, F is the field strength,  $\La$  is a similarity-invariant
 function, $d\mu$ is the invariant measure of the surface;
  $d\mu :={1 \over 2}\epsilon_{\mu \nu} dx^{\mu}dx^{\nu}$.
It was further shown that the partition
function for this theory on a closed non-orientable  manifold,
$\Sigma_{g,s,r}$, with area $A$,  genus $g$, $s$ copies of Klein bottle,
and $r$ copies of projective plane ($RP^2$),  is given by the exact
formula \c{{kh1},{kh2}} as:
 \be\l{2}
 Z_{\Sigma_{g,s,r}}(A)= \sum_{R=\bar{R}} d^{2-(2g +2s +r)}\exp{\br\{}\omega
[-A\Lambda(R)]{\bl\}},
 \ee
where  $R$'s label the irreducible representation of the gauge
group,  which the sum is only over self-conjugate representation,
and also $\La(R)$ is:
 \be \l{3}
  \Lambda (R) =
\sum_{k=1}^p\frac{\alpha_k}{N^{k-1}}C_k(R). \ee
 Here $C_k$ is the k'th Casimir of gauge group, $\alpha_k$'s are
 arbitrary constant. The representation of the $U(N)$ gauge  group
 are labelled by $N$ integers $n_i$ satisfying $n_i \geq n_j$ ($i\leq j$)
 and it is found that:
\bea \l{4}
 d_R&=& \prod_{1 \leq i \leq j \leq N} (1+\frac{n_i-n_j}{j-i}),\\
 C_k(R) &=& \sum_{i=1}^{N}[(n_i+N-i)^k - (N-i)^k].
 \eea
 One defines a function $V$ by
 \be\l{6}
 -N^2V{\br [}A\sum_{k=1}^p \alpha_k {\hat C_k}(R){\bl ]}
  := \omega [-A\Lambda(R)],
\ee where \be\l{7}
 {\hat C_k}(R) = \frac{1}{N^{k+1}}\sum_{i=1}^N (n_i+N-i)^k.
 \ee
 At the large-N limit of the gauge group,
 I use the  following  definitions \c{b3}
\begin{equation}\l{8}
\phi(x) =\lim_{N\rightarrow \infty}\frac{1}{N}(i
-n_i-\frac{N}{2}),
\end{equation}
 where
\be\l{9}
 0\leq x:=\frac{i}{N}\leq 1 \hspace{2cm} {\rm and} \hspace{2cm}
n(x):=\frac{n_i}{N}.
 \ee
Then, apart some unimportant
constant,  the partition function  takes the form:
\begin{equation}\l{13}
Z_{\Sigma_{g,s,r}}(A) = \int^{'} D\phi (x) \exp{\{-N^2S(\phi)\}},
\end{equation}
where
\begin{equation}\l{14}
S[\phi ] = V{\Biggr (}A \int _{0}^{1} W[\phi (x)] dx{\Biggl )}
- (1-(g +s +r/2))\int_{0}^{1} dx \int_0^{1} dy \log|\phi(x)- \phi(y)|,
\end{equation}
and
 \be\l{15}
   W[\phi ] := \sum_{k=1} (-1)^k \alpha_k \phi^k(x).
\ee Note that the sum in (2)  is only over self-conjugate
representations, which $\int'$ in (10) shows this constraint also. This
requirement in $U(N)$ means  that there is the additional
constraint   to the sum as:
 \be\l{10}
 n_i = - n_{N-i +1}.
 \ee
 In the large-N limit, this implies that the continuum variables,
 $\phi(x)$, satisfy:
 \be\l{11}
 \phi(x) = -\phi(1-x).
 \ee
 So one can  define a new function such as:
\be
\phi(x)=\left\{ \bear{cc}
\psi(x) \;\;\;\;\;\;\;\;\;\;\;\;\;\;\;\;\; 0 \leq x \leq 1/2 \\
-\psi(1-x) \;\;\;\;\;\;\;\;\;\;\;\; 1/2 \leq x \leq 1 \eear\right.
\ee Here the function $\psi(x)$ being defined on the interval [0,
1/2], in which $\psi(1/2) = 0$. Then, by institute (15) in (11), it
is seen that these models have interesting solution,  only for
even values of $k$ and therefore one can arrive at:
\begin{equation}
S[\psi ] = V{\Biggr (}2A \int _{0}^{1/2} W[\psi (x)] dx{\Biggl )}
-2(1-(g +s +r/2))\int_{0}^{1/2} dx \int_0^{1/2} dy \log|\psi^2
(x)- \psi^2 (y)|.
\end{equation}
Introducing the density function as $ u(\psi )
:=\frac{dx(\psi)}{d\psi }$ \c{do}. Thereof $W[\psi]$ is an
even function of $\psi$, thus $u[\psi]$ is even, then the
interval corresponding to values of $Z[\psi]$ ($u(\psi)$) is [$-a, a $].
 It is seen that the condition
$n_i\geq n_j$ demands $u(\psi)\leq 1$, and also
 \be\l{16}
 \int_{-a}^{a} u(z) dz = 1.
  \ee
The saddle point that dominates this path integral is given by the
equation of motion. It is found that:
 \bea \l{17}
  h(z)& = &P\int_{-a}^{a}
  \frac{z u(z')dz'}{z^2 - z'^2}\nonumber,\\
&=&P\int_{-a}^{a }\frac{u(z')dz'}{z - z'},
 \eea
where $P$ indicates  the principal value of integral and
 \bea\l{19}
 h(z)&=& \eta W'(z)\\
 \eta&=& \frac{A}{4(1-g-s-r/2)}V'{\biggr\{}A \int_0^{1/2}W[\psi (x)]dx {\biggl \}}, \nonumber \\
&=&\frac{A}{4(1-g-s-r/2)} V'{\biggr\{}A \int_{-a}^a W(z)u(z)dz{\biggl
\}}.
 \eea

 A part from some coefficient, this equation is the same as that
 obtained in \c{{kh1}, {kh2}}, which obtained for this theory on
 orientable surface, and can be solved in the same manner.  the
 density function, $u(z)$, found from (18) depend on the
 modified area $\eta$ and  therefore  $A$, in which as $A$
 increases, a situation is encountered where $u$  exceeds 1. So
 the solution of (18) is  valid only for $A$ less than  some
 critical value $A_c$. $A_c$ is the value of $A$ at which the
 maximum of $u$ becomes 1, $u_{max} (A_c) = 1$. The region
 $A<A_c$ is called the weak coupling phase (WCP) regime  and the
 region $A>A_c$ is called the strong coupling phase (SCP) regime.
 By the same procedure which used in \c{k4}, I can expand the
 density  function in WCP regime, $u_w(z)$, around absolute
 maxima, $z_0$, and it is found that for the points which are
 adjacent of critical point, $A_c$, the difference of free
 energy in SCP and WCP regime is:
 \be\l{20}
 F_s-F_w \simeq \xi^3,
  \ee
  where
 the free energy of the theory is defined as:
  \be\l{21}
 F := S|_{\phi_{cla.}},
 \ee
 and
 \be\l{22}
 \xi = u_w(z_0) - 1.
 \ee
 By considering $\xi$ as a function of $A$ and expand it around $A = A_c$,
 one can arrive at:
 \be\l{23}
 \xi(A) = \xi'(A)(A-A_c) + \ldots
 \ee
 where
 \be\l{24}
 \xi' ={\Br (} \frac{\partial \xi}{\partial \eta}{\Bl )}_{\eta = \eta_c}
 \frac{d\eta}{dA}|_{A=A_c}.
 \ee
So for the case which $\frac{\partial \xi}{\partial \eta}|_{\eta=\eta_c}$
and $\frac{d\eta}{dA}|_{A=A_c}$ are nonzero, we have
 \be\l{25}
 F_s-F_w \simeq \beta (A-A_c)^3 + \ldots
   \ee
 Here $\beta$ is a constant which is independent of modified area
 of manifold, $\eta$ (or $A$). Thus, almost all models which, the density
 function of that have some absolute maxima, and exist the proper
 quantity $\xi(A)$ in which, $\frac{d\xi}{dA}\neq 0$,
 has a third order phase transition on any arbitrary non-orientable
 surface.

\section{$W(\phi) = \phi^{2k} (k>1)$ models.}

In this case $W(\phi)$ is an even function of $\phi$,
 then we can use of (18). So by solving (\r{17}), it is found that:
\be\l{26}
 u_w(z) ={k\eta \over \pi}\sqrt{a^2-z^2}\sum_{n=0}^{\infty}
 \frac{(2n-1)!!}{2^nn!}a_k^{2n}z^{2k-2n-2},
\ee where $a_k$ is obtained from (\r{16}) as:
 \be\l{27}
 a_k ={\Br[}\frac{2^k(k-1)!!}{(2k-1)!!\eta}{\Bl ]}^{\frac{1}{2k}}.
\ee
 It has been shown that  (\r{26}) has three extremum points at $z=0$  and
$z_{1,2} = \pm \alpha_k\sqrt{\zeta_k}$ \c{Ali}, in which $\zeta_k$ is
independent of $a_k$ and is determined from
 \be\l{28}
\sum_{n=0}^{k-2}\frac{(2n-1)!!}{2^{n+1}(n+1)!}\zeta_k^{-(n+1)} =
1. \ee Using $u'_w(z_k) = 0$, (where $z_k$'s are extremum points),
one can see:
 \be\l{29}
u''_w(z_0) =-\frac{k\eta
a^{2k-2}(2k-1)}{\pi\sqrt{a^2-z_0^2}}\sum_{n=0}^{k-2}
\frac{(2n-1)!!z_0^{2k-2n-4}}{2^nn!a^{2k-2n-4}}.
 \ee
 So when $z_k = z_0 =0$, then $u''_w(z_0) =0$, so that for
 $\psi^{2k}$($\phi^{2k}$) models the density function $u_w(z)$ is minimum at
 $z_0 = 0$ but for $z_k = z_{1,2}$ the density function is maximum
 only  for cases which $\eta >0$ or
 \be\l{30}
 1> g + s + r/2,
 \ee
 therefore , we see that the density function of these theories is
 maxima at $z_k = z_{1,2}$, when (\r{30}) has been satisfied.
By substitute (\r{27}) in (\r{26}), we obtain:
 \be\l{31}
u_{w}(z_{1,2})=\eta^{\frac{1}{2k}}f(k)
 \ee
where $f(k)$ is independent of $\eta$(or $A$). Using (24) and
(25), we have
 \be\l{32}
  \xi(A) =
{\br \{}\frac{d}{dA}\ln\eta^{\frac{1}{2k}}{\bl \}}_{A=A_c}(A-A_c)
+ \ldots \ee where, we use of $\xi(A_c) = 0$. Thus if
$\frac{d\ln{\eta^{\frac{1}{2k}}}}{dA}|_{A=A_c}$ is nonzero, these
theories has a third order phase transition. Note that for the
case which $g=s=r=0$, the relation (\r{30}) is satisfied and this
means that these theories has third order phase transition on
sphere, also if $g=s=0$ and $r=1$, the order of phase transition
of theory on projective plane (RP$^2$) is 3 for all order of
$\psi^{2k}$, but if $g=r=0$ and  $s=1$,   the condition (\r{30})
is not satisfied, then the theory has no phase transition on Klien
bottle and other non-orientable surfaces.

\section{the $\psi^2 + \frac{\gamma}{4}\psi^4$ models.}

I now consider $\psi^2 + \frac{\gamma}{4}\psi^4$ with arbitrary
$\gamma$. At the first I assume that  $\gamma > 0$.
So by solving (\r{17}), one can obtain:
\be\l{33}
u_w(z) = \frac{\eta}{4\pi}\sqrt{a^2 - z^2}(4 + \gamma
a^2 + 2\gamma z^2),
 \ee
and from (\r{16}),
 \be\l{34}
\eta a^2(8 +3\gamma a^2) = 16.
 \ee
 The shape of $u_w(z)$ depends on $\eta$.
   It is seen that
\be\l{35}
 u''_w(0) =\frac{\eta}{4\pi}(3\gamma a^2 -4).
  \ee
  So for  $3\gamma a^2 < 4$,  $u_w(z)$ has only one extremum  point at
 $z=0$, if $\eta > 0$, and also
 \bea\l{36}
 \xi'(A_c) &= &\frac{d(u_w(0) -1)}{dA}|_{A=A_c}, \nonumber \\
 &=& {1 \over \sqrt{108 \pi^2\gamma}}{\br \{}
 {12\gamma / \eta_c + \sqrt{1 + 48\gamma / \eta_c} - 1 \over
 \sqrt{\sqrt{1 + 48\gamma /\eta_c} -1}}{\bl \}}.
 \eea
Thus,  the  density function is maximum at $z=0$ ($u''_w(0) <0$)
for the case which $\eta >0$ ($1-g-s-r/2 >0$), and this is
satisfied only on sphere and projective plane. However the order
of phase transition of this model is three on sphere and
projective plain and on other orientable and non-orientable
surface has no phase transition. For $3\gamma a^2 > 4 $  this
model has three extremum point at $z=0$ and $z=\pm z_0$, where
\be\l{37} z_0= \sqrt{\frac{2}{3\gamma}{\br (}\sqrt{1+
\frac{48\gamma}{\eta}}-2} {\bl )}. \ee From this for the case
which $\eta >0$, one can obtain that $u_w(z_0)$ is minimum at
$z=0$ and has two maxima at $\pm z_0$. So by obtaining $u_w(z_0)$
and use of (23-25) , one can arrive at:
 \be\l{38} \xi'(A_c) =
\sqrt{\frac{2}{3\gamma}}\times \frac{4\eta_c + 3\gamma}
{6\pi\eta^{3/4}(\eta_c +3\gamma)^{1/4}}\times
\frac{d\eta}{dA}|_{A=A_c}
>0.
\ee
So according with (\r{24}) and (\r{25}), this theory has phase transition only
on S$^2$ and RP$^2$ and the order of it is 3.
For the case which $\gamma <0$(for all value of $\gamma$),
 the density function $u_w(z)$ has
three extremum point at $z=0$ and $\pm z_0$,  where
\be\l{39}
z_0 = \sqrt{\frac{2}{3|\gamma|}{\br (}2 + \sqrt{1 +
3|\frac{\gamma}{\eta}|}{\bl )}}.
\ee
It is clearly seen that this model of theory has phase transition
 on all surfaces which $\eta <0$. Indeed, it  is not difficult to check
 that $u''_w(z_0)$ is negative, and also, from (\r{39}), one can obtain
 $u_w(z_0)$, and then:
 \bea\l{40}
 \frac{d\xi}{dA }|_{A=A_c}& = & \frac{d}{dA}(u_w(z_0) - 1)|_{A=A_c},
 \nonumber \\
 &=& \frac{2}{3\pi}\sqrt{\frac{2}{3|\gamma|}}\times
 \frac{d\eta}{dA}|_{A=A_c},
 \eea
 so according with (\r{24}) and (\r{25}), it is found that if
  $\frac{d\eta}{dA }|_{A=A_c} \neq 0$, then  the $\psi^2 + \frac{\gamma}{4}
  \psi^4$ model whit $\gamma<0$ has a third order phase
  transition on all closed orientable  and non-orientable
  manifold except on sphere, torus, projective plain and Klein bottle.

\section{conclusion}

I study the large-N limit of  nlgYM$_2$ theories on  arbitrary
non-orientable surface. By obtaining the effective action of these
theories at the large-N limit, it is shown that apart some
coefficient, the saddle point equation is the same as that obtain
for these theories on orientable surface. I study the phase
structure of
 $\phi^{2k}$ model
 and it is seen that all order of this model has
 third order phase transition only on projective plane.
 Also, by considering   the $\phi^2 +
\frac{\gamma}{4}\phi^4$ model of  nlgYM$_2$, I found that, for the
case $\gamma > 0 $, this theory has third order phase transition
on $\bf RP^2$, and on other non-orientable  surface has no phase
transition, whereas for the case $\gamma <0$ this model has phase
transition on all closed non-orientable surfaces except projective
plane and Klein bottle, and the order of phase transition is 3. It
is clearly seen that all order of $\phi^2 + \alpha \phi^{2k +1}$
model on any arbitrary closed non-orientable surfaces has the same
phase structure of $\phi^2$ model of nlYM$_2$.\\
Note that the whole reasoning is independent  of the number of
points
 as $u_w$ attains its absolute maximum. It is clear that a similar
 situation prevails for the cases which $u_w$ has many absolute
 maximum, and  also in this case, one can realize the WCP regime
 and obtain the phase structure of theory in the multi-critical
  points. It is remarkable that there might be cases which
   $u''_w(z_0)$ is zero (not negative). At this state the theory
   has phase transition, but the order of phase transition is not
   3.

\end{document}